\documentclass[twocolumn,showpacs,preprintnumbers,amsmath,amssymb,superscriptaddress]{revtex4}
\usepackage{graphicx}
\usepackage{dcolumn}
\usepackage{bm}
\graphicspath{{./Figures/}}

\begin{document}

\preprint{APS/123-QED}

\title{Probing space charge and resolving overlimiting current mechanisms at the micro-nanochannel interface}

\author{Jarrod Schiffbauer}%
\author{Uri Liel}%
\author{Neta Leibowitz}%
\author{Sinwook Park}%
\author{Gilad Yossifon}%
\affiliation{%
Faculty of Mechanical Engineering, Micro- and Nanofluidics Laboratory, Technion - Israel Institute
of Technology - Technion City 32000, Israel
}%

\date{\today}

\begin{abstract}
We present results demonstrating the space charge-mediated transition between classical, diffusion-limited current and surface-conduction dominant over-limiting current in a shallow micro-nanochannel device. The extended space charge layer develops at the depleted micro-nanochannel entrance at high current and is correlated with a distinctive maximum in the dc resistance. Experimental results for a shallow surface-conduction dominated system are compared with theoretical models, allowing estimates of the effective surface charge at high voltage to be obtained. In comparison to an equilibrium estimate of the surface charge obtained from electrochemical impedance spectroscopy, it is further observed that the effective surface charge appears to change under applied voltage.
\end{abstract}

\maketitle
\begin{figure}
  \includegraphics[width=3.4 in]{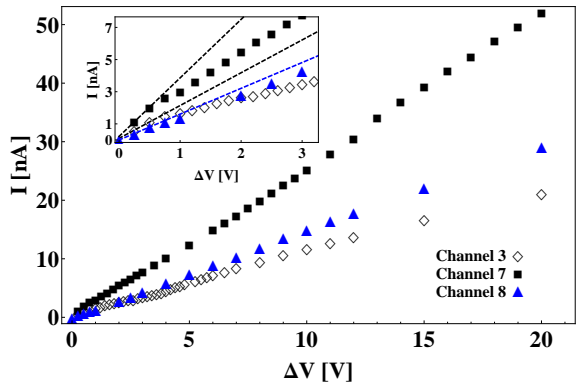}\\
  \caption{(Color online) Current-voltage characteristics for several 3 $\mu$m-deep micro-nanochannel devices. The inset shows the transition from diffusion-limited, pseudo-Ohmic response to OLC, estimated by the point of departure from the linear fit (dashed lines). For all three channels, this is between 0.75 V and 1 V, with the transition for channel 7 slightly lower than the others. Channels 3 and 7 taken with Pt electrodes, channel 8 with Ag/AgCl.}\label{fig:f2}
\end{figure}
\begin{figure}
  \begin{center}
  \includegraphics[width=3.6 in]{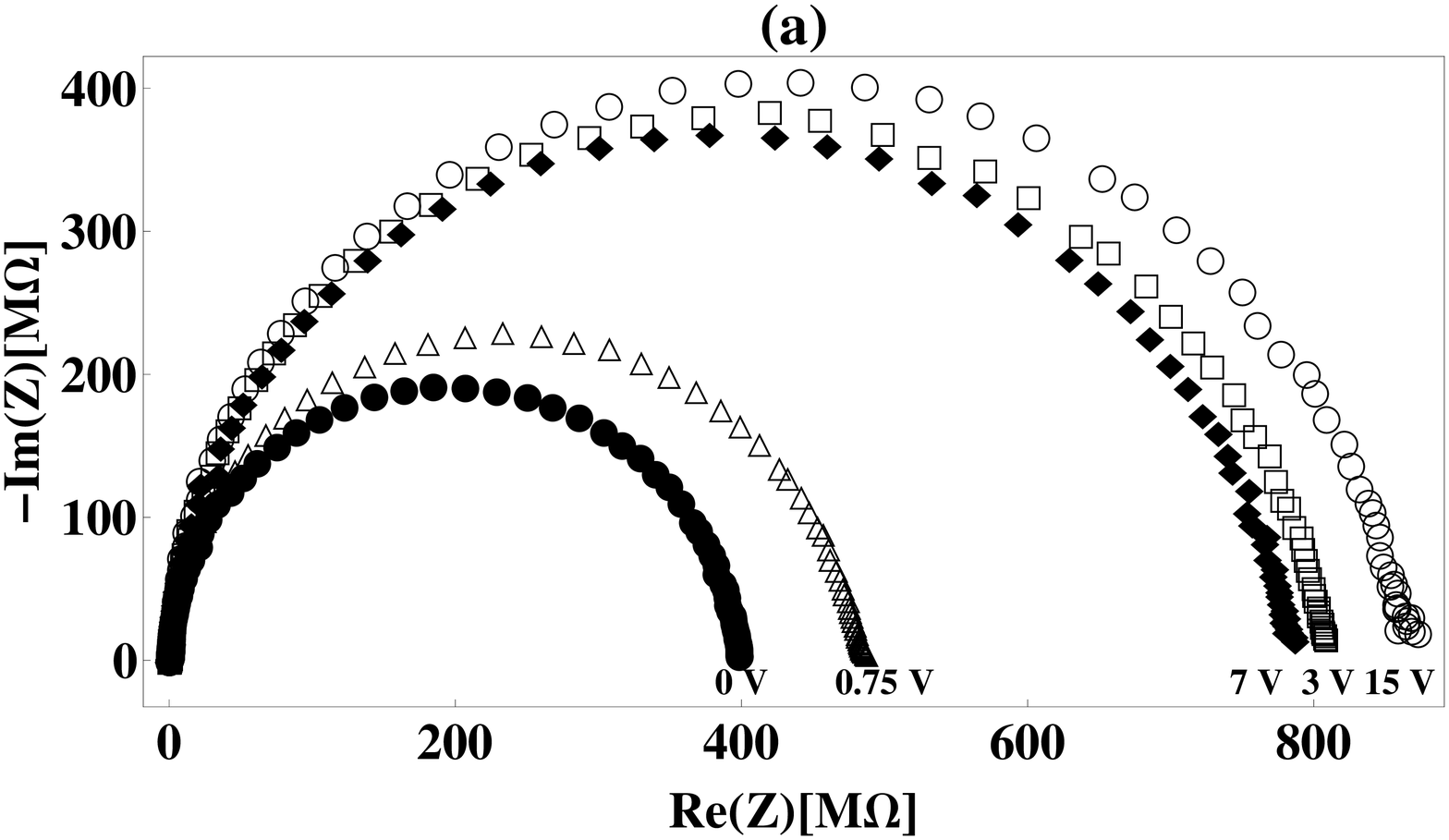}\\
   \includegraphics[width=3.6 in]{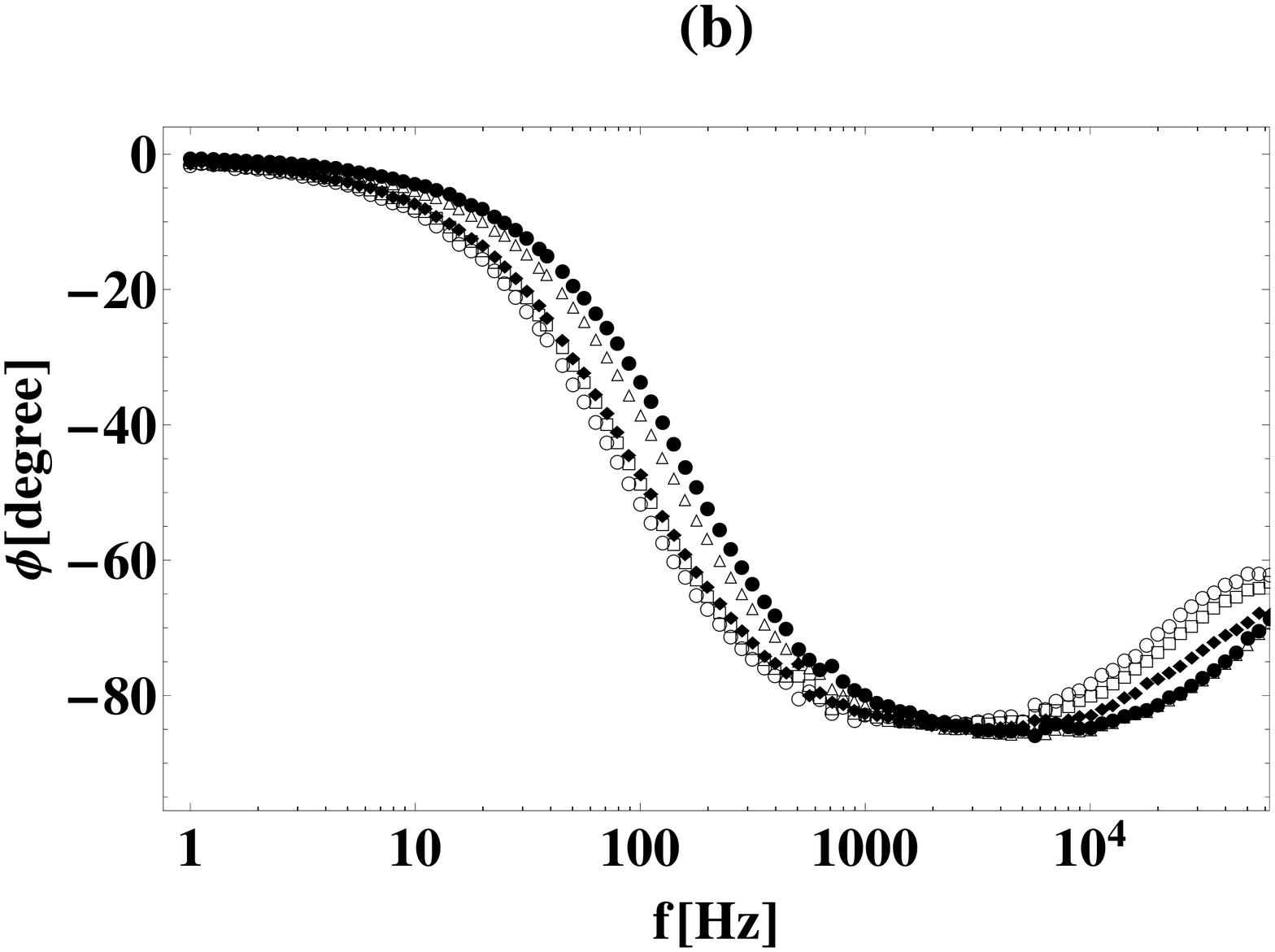}\\
   \end{center}
  \caption{Nyquist (a) and Bode phase (b) plots for channel 3 showing typical EIS response across a range of dc bias voltages. Data for other channels are shown in Figs. ~\ref{fig:f7} and~\ref{fig:f8} in the appendix}\label{fig:f3}
\end{figure}
\begin{figure}
  \includegraphics[width=3.6 in]{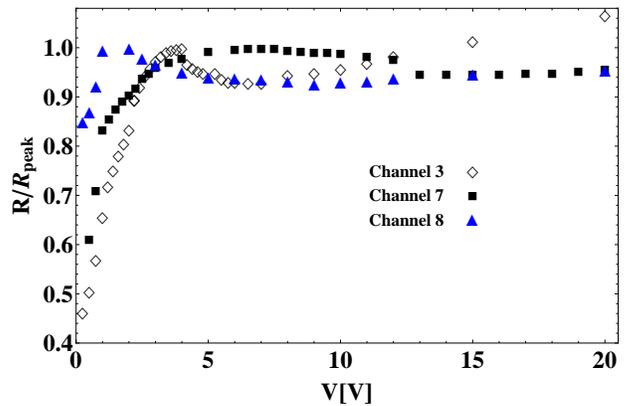}\\
  \caption{(Color online) Experiment R-V plots (static R) obtained from the slow I-V curve, normalized by peak resistance. Channels 3 and 7 taken with Pt electrodes, channel 8 with Ag/AgCl.}\label{fig:f4}
\end{figure}
\begin{figure}
   \begin{center}
  \includegraphics[width=3.6 in]{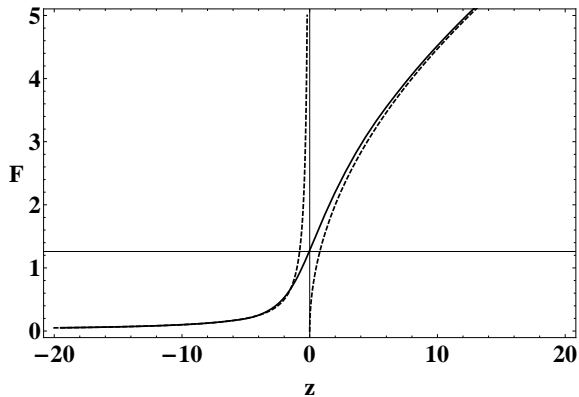}\\
  \end{center}
  \caption{Numerical Painleve solution for F(0) (solid) and wide space charge asymptotes (dashed).}\label{fig:f5}
\end{figure}
\begin{figure}
   \begin{center}
  \includegraphics[width=3.6 in]{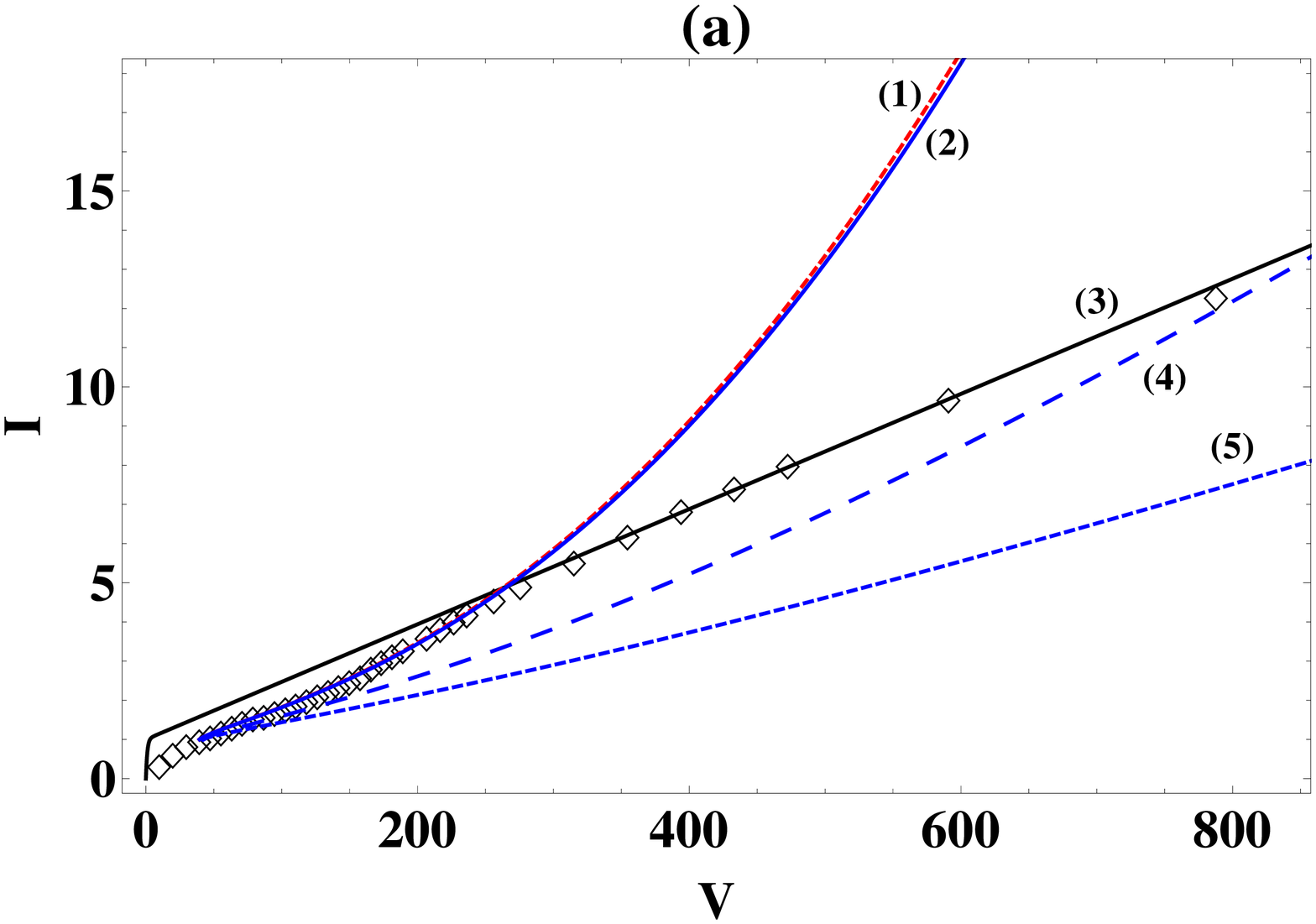}\\
  \includegraphics[width=3.6 in]{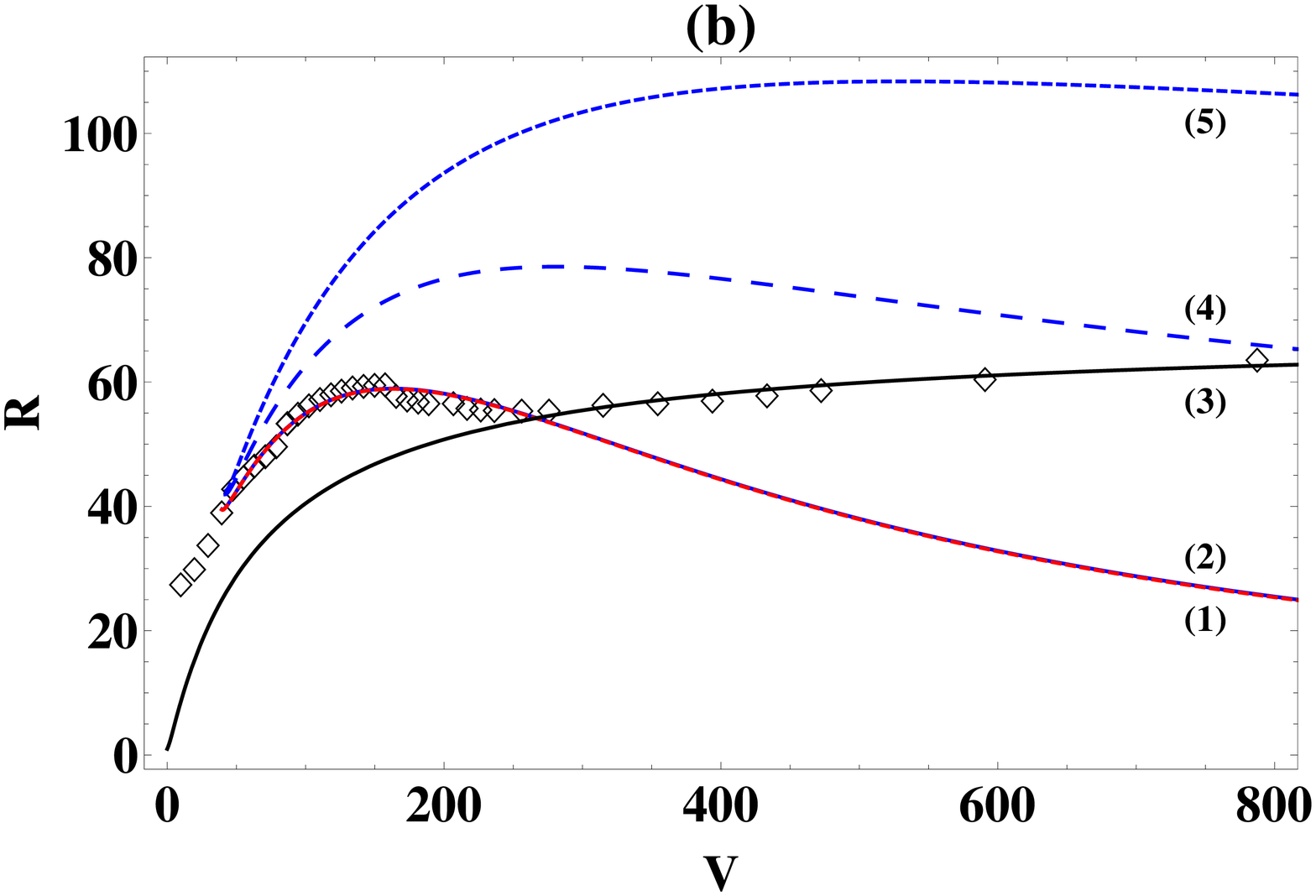}\\
  \end{center}
  \caption{(Color online) Current-voltage response (a) and static resistance, $R=V/I$, (b) for theoretical models and channel 3 (hollow diamonds), normalized to apparent limiting current ($\sim 1.8$ nA) and thermal voltage $V_{th}=0.0254$ V. (1) ESC-only model (red dashed) with $\epsilon=0.0067$ and $a=0.63$, used for all models involving the ESC, (2) ESC+DL$||$SC model (solid blue) with $\rho_s=-18.49$, corresponding to the microchannel equilibrium surface charge density for channel 3, (3) the DL$||$SC model of Dydek et al~\cite{DydekPRL} with $\rho_s=-0.0147$, (4) ESC+DL$||$SC model with $\rho_s=-0.0347$ and (5) ESC+DL$||$SC with $\rho_s=-0.0147$}\label{fig:f6}
\end{figure}

The non-linear electrochemical response of ion-selective interfaces, e.g, electrodes, electro-deposition, shock electrodyalsis, ion exchange membranes, or fabricated nanochannels, has been the subject of a tremendous amount of research~\cite{RZN,RZbias,Sistat2008,Nikonenko,Schoch,DSteinPRL,HeydenPRL,KimPRL,Nanoreview,Naturenano,RUBSHTILL79,RZPRE2000,JFM2007,RZPRL,YCPRL1,DydekPRL,KimBazant,Khoo,DengSuss}. Much of this work concerns how the dc current-voltage response emerges from a variety of mechanisms~\cite{KimPRL,Nanoreview,Naturenano,RUBSHTILL79,RZPRE2000,JFM2007,RZPRL,YCPRL1,DydekPRL,KimBazant}. At low voltage, the response appears Ohmic as concentration polarization (CP) develops across the system resulting in depletion and enrichment of ions at opposite interfaces. At sufficiently high (limiting) current~\cite{Levich}, the concentration approaches zero at the depleted interface. Electroneutrality cannot be maintained and a distinct non-equilibrium extended space charge (ESC) develops~\cite{RUBSHTILL79} in this region. This transition from Ohmic response is identified with a decrease of the I-V slope, corresponding to the increased resistance of the system due to depletion and subsequent ESC development (see for example, Fig.~\ref{fig:f2}.) Often, there is another distinct transition wherein the slope increases again. Herein we shall refer to all currents above the first transition, corresponding to the classical limiting regime, as over-limiting currents (OLC).\\
\begin{figure}
  \includegraphics[width=3.4 in]{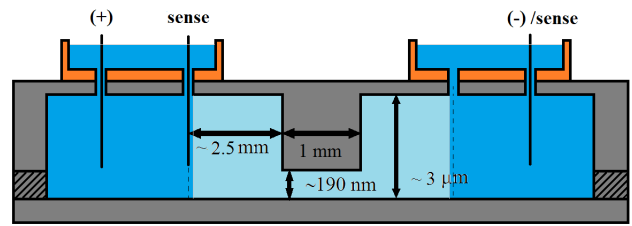}\\
  \caption{(Color online) Schematic device diagram showing device dimensions and electrode placement, where the sensing electrode on the anodic side is separate from the biased electrode (the same electrode is used on the cathodic side for both sensing and bias.) Notice the maximum possible extent of the diffusion layer is fixed by the position of the first drill hole enforcing bulk-like concentration, denoted by dark-light shading. The width (normal into the page) of all channels is approximately 2 mm.}\label{fig:f1}
\end{figure}
\indent Several mechanisms contribute to or affect the OLC~\cite{RUBSHTILL79, RZPRE2000, JFM2007, RZPRL, YCPRL1,DydekPRL,KimBazant,Khoo,DengSuss,ManiB, Yaroschuk, NBruus, Khair, ChiaEugene}. Which mechanisms dominate the response depend on the applied conditions and details of the particular system. Dydek et. al proposed a model for microchannel devices which predicts the dominant mode of OLC based on microchannel depth and surface charge~\cite{DydekPRL}. The shallowest channels are dominated by surface conduction~\cite{ManiB,DengSuss,KimBazant}, intermediate depth channels by non-uniform electro-osmotic flow (EOF) effects~\cite{Yaroschuk, NBruus}, and deep channels, or unconfined systems by electrokinetic instability~\cite{RZPRE2000, JFM2007, RZPRL, YCPRL1}. The earliest proposed mechanism for sustaining OLC is the existence and structure of the ESC itself~\cite{RUBSHTILL79}. However, while the role of surface conduction in the OLC of shallow channels at high voltage has been confirmed experimentally~\cite{KimBazant}, the direct role played by the ESC in the OLC response has received little attention. Consequently, how it interacts with the surface conduction mechanism, specifically in mediating the transition between classical diffusion-limited currents and the high-voltage OLC is not understood.\\
\indent Thus the goal of the present paper is to better understand the interaction between the ESC and surface conduction in the over-limiting response of a shallow micro-nanochannel device. Hence, the present work extends and generalizes previous studies~\cite{DydekPRL,KimBazant}. The device, shown schematically in Fig.~\ref{fig:f1}, has 3 $\mu$m deep microchannels, so the response will be dominated by electro-diffusive transport with convection being of secondary importance~\cite{DydekPRL,Yaroschuk}. Fabrication of the Pyrex-silicon devices are similar to those used in our previous studies and discussed in detail elsewhere~\cite{ImpPRL}. The results here were obtained using a KCl electrolyte at concentrations on the order of $10^{-5}$ M to ensure high nanochannel selectivity, with a (bulk) pH of about $5.6$. Platinum electrodes were used in a 3-electrode set-up to help isolate the effects of electrode reactions from the measurement. Additional data were taken using Ag/AgCl electrodes to verify that the effects of electrolysis of water are not of critical importance to the phenomena of interest.\\
\indent Each channel was wetted with KCl for 16-24 hours (overnight) prior to conducting the experiment to ensure proper wetting without introduction of air bubbles into the channels. For channel cleaning, a 10 V dc voltage is applied for 500 seconds in each direction until current was approximately the same magnitude at end of each cycle. This typically took 4-6 cycles. Reservoirs were flushed and filled with fresh electrolyte at end of each cycle. The system was allowed to relax back to (approximate) zero-current state prior to running a pre-conditioning cyclic voltammetry (CV). The CV runs from 0 V to 20 V to -20 V back to 0 V in 100 mV steps with sweep rate of 20 sec per step, or 5 mV/sec. From previous experience, the system conductivity can fluctuate somewhat during first few hours of operation, possibly due to flow-induced dissolution-based changes in surface charge~\cite{MBonn}. The current is run in both directions to avoid asymmetric changes in the channel. After this, the system is relaxed back to an approximate zero-current state prior to actual data collection.\\
\indent Electrochemical impedance spectroscopy (EIS) is used to characterize the system. Measurements are made using a Gamry 3000 potentiostat with 50 mV perturbation voltage. EIS spectra are taken from 1 MHz down to about 1 Hz~\footnote{The distances to the bulk reservoir ports, which determines roughly the electroneutral diffusion time and hence the frequency for the maximum of the (microchannel) Warburg EIS response, is approximately 2 mm. For a KCl electrolyte, we expect this frequency to be in the mHz range. Thus no appreciable Warburg behavior appears in the Nyquist plots}. For the EIS, we used a 3-electrode set-up for consistent operation in the high-voltage mode (HVE) and some isolation of the sensing electrode from possible reactions when under dc bias. EIS taken at 0 V dc bias is used to obtain the equilibrium resistance of the device, $R_o$, from fitting with a simple parallel RC circuit. While this omits some important details which can be obtained from a more physically realistic fundamental transport model, the equilibrium resistance value is reasonably reliable~\cite{surfchgPRE,EDLthryPRE}. The system is then subject to dc voltage steps from 0.25 V to over 20 V. A 300-second conditioning voltage is applied at each step to allow development of concentration polarization (CP), and the current response recorded. A dc-biased EIS measurement is then conducted immediately following the conditioning step. An I-V curve is obtained from the current at the end of the conditioning step (Fig.~\ref{fig:f2}), corresponding to a relatively well-developed CP and response very close to that of the steady-state.\\
\indent The equilibrium resistance, $R_o$, can be used to estimate the surface charge at the outset of the experiment. Both microchannels and the nanochannel are assumed to be in thermodynamic equilibrium with the reservoirs, the surface charge density is assumed uniform, cross-sectional electro-neutrality is assumed to hold throughout the device, and fluid flow/streaming effects are neglected. Such an approach has been shown to yield reasonable values for the surface charge across a range of concentrations, in comparison to more detailed approaches~\cite{surfchgPRE,EDLthryPRE}. Using cross-sectionally averaged electrochemical potentials, the effective conductivity (in S m$^{-1}$) of a channel segment with a symmetric, monovalent electrolyte of bulk concentration $c_o$ (in mol m$^{-3}$) and mobility $\mu$, is given by $\Sigma=2 F\mu c_o \sqrt{\rho_s^{2} + 1}$. $F$ is the Faraday constant and $\rho_s$ (dimensionless with scaling $2Fc_o$) has the same meaning as in Ref.~\cite{DydekPRL}, an effective volumetric charge density related to the surface charge, $\sigma_s= \rho_s F h c_o$ (in C m$^{-2}$) with channel height $h$. By using this effective conductivity, the surface charge density can be obtained from the total resistance, $R_{tot}=2L_{m}/A_{m}\Sigma_{m} + L_n/A_n\Sigma_n$, where the subscripts ``$m$" and ``n" denote micro- and nanochannel segments respectively. For channel 3 with $R_o = 395$ M$\Omega$ (see Fig.~\ref{fig:f3}), and $\mu=7.6 \times 10^{-8}$ m$^2$(sV)$^{-1}$ for KCl, this gives $\sigma_s =-0.136 $ C m$^{-2}$, which is somewhat high for this range of electrolyte concentration compared to similar devices~\cite{surfchgPRE} but within reason~\cite{DSteinPRL}.\\
\indent Nyquist and Bode plots for a representative series of EIS on channel 3 are shown in Fig.~\ref{fig:f3}, illustrating the impedance response as a function of dc bias voltage. Both imaginary and real impedance components initially increase through the Ohmic response and into the lower portion of the OLC (compare with I-V in Figs.~\ref{fig:f2} or~\ref{fig:f6}). In many cases, at a given voltage, the impedance reaches a local maximum and is observed to decrease with subsequent increase in voltage. At even higher voltages, there is either a slight rise in or saturation of the impedance. The Bode phase reflects this. The transition from resistive to capacitive response, i.e. from $\phi = 0$ to $\phi=-90$ degrees, occurs at lower frequencies with increasing voltage, corresponding to increasing CP and the formation and growth of the ESC. Similar behavior is observed for the equilibrium impedance of an isolated, thin channel with changing bulk concentration~\cite{surfchgPRE,EDLthryPRE}.\\
\indent This same behavior is seen in the dc resistance obtained from the I-V data. Several such plots of dc resistance vs. voltage are shown in Figure~\ref{fig:f4}. The resistance is normalized by the height of the resistance maximum for clarity on the same plot. The resistance maximum typically occurs above the voltage corresponding to classical Ohmic-to-limiting transition, at about 3 V and 1 V respectively for channel 3.\\
\indent The experimental results may be understood with the help of two simple 1D models for the dimensionless current density-voltage response of a 1-layer ion-selective systems, one representing the ESC response, and the other electro-diffusion modified by surface conduction. The static resistance of the ESC for an ideal membrane,
\begin{equation} R =\frac{ 4(-1-x_o)^{3/2}}{3\epsilon \sqrt{2I}},\end{equation}
can be obtained from the well-known ESC voltage drop and estimates for the ESC outer edge, $x_o=2/I - 3- a(2\epsilon^2/I)^{1/3}$,~\cite{JFM2007}. This resistance yields a prominent maximum shortly above the limiting transition, as seen in Fig.~\ref{fig:f4}. In the ESC model, the parameter $\epsilon =\lambda_D/L_o$ is the dimensionless Debye length and the parameter $a = 0.63$ is obtained from a numerical solution of the Painleve problem, shown in Fig.~\ref{fig:f5}. We briefly discuss the most important details below and refer the reader to a more comprehensive discussion elsewhere~\cite{JFM2007}.\\
\indent It is well-known that the 1D electrodiffusive problem can be cast in terms of a master equation for the (scaled) electric field, $E=\epsilon \psi$, where $\psi$ is the electric potential satisfying the Poisson equation. This takes the form,
\begin{equation}\epsilon^2 \frac{d^2 E}{dx^2}-\frac{E^3}{2}+I E (x-x_o)=-\epsilon I \label{eqn:e8a}.\end{equation}
For the present purpose, we require the electroneutral salt concentration to estimate the value of $x_o$, giving a measure of the extent of the space charge. In the full formulation,
\begin{equation}c(x)=\frac{E^2}{4}-\frac{I}{2}(x-x_o) \label{eqn:e8b}\end{equation}
and thus to obtain the parameter estimate, we require $E(x_o)$. By re-scaling the problem, asymptotic solutions valid for fully-developed space charge may be obtained and used as boundary conditions on the numerical solution of the re-scaled problem,
\begin{equation}F''-\frac{1}{2}G^2 F^3 +zF=-1 \label{eqn:e10}\end{equation}
where the transformations
\begin{equation}E= (\epsilon I)^{1/3} F \end{equation}
\begin{equation}(x-x_o) = I^{-1/3} \epsilon^{2/3} z \end{equation}
have been used. The asymptotes for developed space charge are given,
\begin{eqnarray}
F(z) = \left\{
        \begin{array}{ll}
            -\frac{1}{z} & \quad z\ll 0 \nonumber \\
            \quad \quad \nonumber \\
            \sqrt{2z} & \quad z\gg 0
        \end{array}
    \right.
\label{eqn:e11}\end{eqnarray}
and the full numerical solution is shown in Fig.~\ref{fig:f5}.
\indent The diffusion-limited transport plus surface-conduction (DL$||$SC) model from Ref.~\cite{DydekPRL},
 \begin{equation}I = 1-e^{-V}-\rho_s V ,\end{equation}
shows an approximately constant I-V slope at high voltage, similar to experimental data. The low voltage for the Ohmic-to-limiting transition for this model (curve (3) in Fig.~\ref{fig:f6}) is a consequence of an ideally selective interface; lower selectivity tends to shift the transition to higher voltage. The voltage is scaled by the thermal voltage, $RT/F\approx 0.0254$ V, the concentration by the bulk value, and the length by the extent of the (assumed 1D) depletion region, $L_o\sim O(1)$ mm. The actual diffusion length is difficult to determine without taking into account field-focusing and any EOF effects, but this is a reasonable estimate. The current density is scaled by $2c_o\mu RT/L_o$ The Einstein relation is assumed to hold between mobility and diffusivity, and the conductivity assumed due solely to the electrolyte; we do not attempt to account for dissolved CO$_2$, etc. The DL$||$SC model has the effective microchannel charge density, $\rho_s$, as a parameter. Each basic model is fitted to the most appropriate region of the I-V and the combined model, ESC+DL$||$SC, is obtained by placing the two models in series (Fig.~\ref{fig:f6}).\\
\indent Both basic models are derived for a 1-layer ideally selective interface, and do not include the contributions of field-focusing~\cite{YCffPRE} or of the nanochannel itself. Furthermore, the ESC model is not applicable below the classical limiting current. Hence, calculations and plots involving the ESC model are shifted by the Ohmic-to-OLC transition voltage, which in a sense accounts for both non-ideality (up to the limiting current) and the non-zero equilibrium resistance of the nanochannel. While neither model can capture the entire curve, each can fit a respective region of normalized experimental data reasonably well. The fit obtained for the I-V of channel 3 has $\epsilon = 0.0067$ and is quite good around the lower OLC voltages (see Fig.~\ref{fig:f6}.) Fitting the DL$||$SC model to the high-voltage OLC, one obtains $\rho_s = -0.0147$, considerably lower than the value obtained from the equilibrium resistance ($\rho_s=-18.49$ for the microchannel). For the combined model, larger $\rho_s$ corresponds to a smaller voltage drop across the depleted electro-neutral portion of the microchannel, a larger drop across the non-electroneutral ESC, and a more pronounced maximum (see Fig.~\ref{fig:f6}). For $\rho_s=-18.49$, the combined model agrees quite well up to the maximum, where the models with low surface charge begin to agree better with data.\\
\indent Analyzing the behavior of these simple models leads to a number of general conclusions. The character of I-V response between the transition from diffusion-limited current and surface-conduction dominant OLC is governed primarily by the ESC at intermediate voltages, provided the other resistances in the system are not too large by comparison. As can be seen in Fig.~\ref{fig:f6}, case (2), at sufficiently high voltage, the enhanced electric fields within the ESC can actually drive the resistance below the limiting value. Of course, it should be clear that this is only the case when the other resistances in the system are negligible (or themselves decrease.) For the DL$||$SC model with constant and uniform surface charge, the differential and static resistance have the common high-voltage limit, $-1/\rho_s$. Hence, in the combined model the surface charge, through $\rho_s$, determines the high-voltage resistance and affects the shape and height of the observed resistance maximum in conjunction with $\epsilon$.\\
\indent In the case of low surface charge, the corresponding voltage drop can become large. So at a given over-limiting current, the voltage drop across the ESC is smaller for lower surface charge. This results in a broader, flatter maximum, shifted to higher voltages and with overall higher resistance. For low enough surface charge, the corresponding high resistance renders the ESC essentially unobservable (see case (5) in Fig.~\ref{fig:f6}).\\
\indent The exceedingly large difference between equilibrium and high-voltage estimates for the effective surface charge is also very interesting. The simplifications made in the modeling approach have omitted a number of things, e.g. the explicit consideration of CP in the enriched microchannel, non-ideality of the nanochannel (nanochannel CP)~\cite{AbuRjal}, field-focusing~\cite{YCffPRE,YoavShachar}, and fluid-flow~\cite{HeydenPRL,ImpPRL}. While these can be expected to contribute somewhat, it is
nonetheless difficult not to conclude that the assumption of spatially uniform surface charge, independent of applied voltage, is itself suspect. In light of studies on surface charge regulation via silanol dissociation~\cite{BehrensGrier} and determination of a concentration-dependent effective surface charge in equilibrium nanochannels~\cite{surfchgPRE}, concentration polarization itself should lead to both non-uniformity and voltage-dependence of surface charge. Of course, this is also tied strongly to the selectivity of the nanochannel, which not only affects the I-V response in the classical under-limiting regime, but also can be expected to play a role in the structure of the ESC. The field-focusing effect of the 2D interface is known to have a strong effect on the CP profile and voltage-drops across the system~\cite{YCffPRE,YoavShachar}, and therefore may be anticipated to affect the ESC as well. Thus, the somewhat high fitted value of $\epsilon$ can at least partially be attributed to this. Lastly, it has also been shown that there is an additional current path for surface conduction in parallel to the diffuse EDL, which may be associated with Stern layer transport~\cite{EDLthryPRE}. Future and on-going work concerns extending the models to capture non-ideality and field-focusing, understanding the apparent change in surface charge between equilibrium and high-voltage, and extending the experiments to deeper channels where EOF is expected to play an important role.\\
\begin{acknowledgments}
The authors acknowledge the Israel Science Foundation for grant number 2015240, the Technion Russel-Berrie Nanotechnology Institute (RBNI), and Stephen and Nancy Grand Water Research Institute for grant number 2017720, for financial support and the Technion Micro-Nano Fabrication Unit (MNFU).\\
\end{acknowledgments}
\appendix*
\section{Additional Experimental Data}
Additional Nyquist and phase plots for channels 7 and 8 are shown below. This compliments the data shown in Fig.~\ref{fig:f3} for channel 3. The normalization used in Fig.~\ref{fig:f4} can be inferred from the $Re(Z)$ intercepts. 
\begin{figure}
   \begin{center}
  \includegraphics[width=3.6 in]{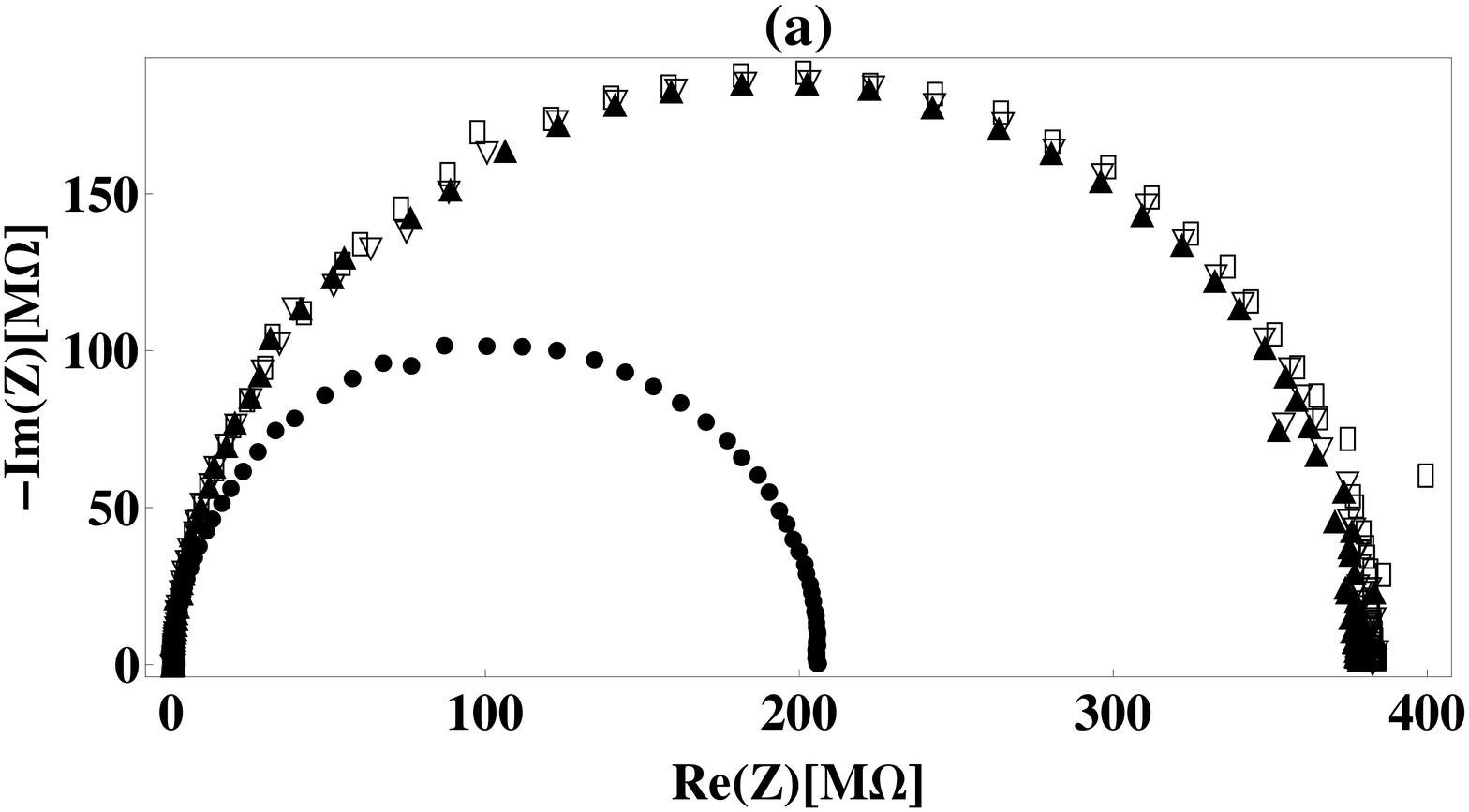}\\
  \includegraphics[width=3.6 in]{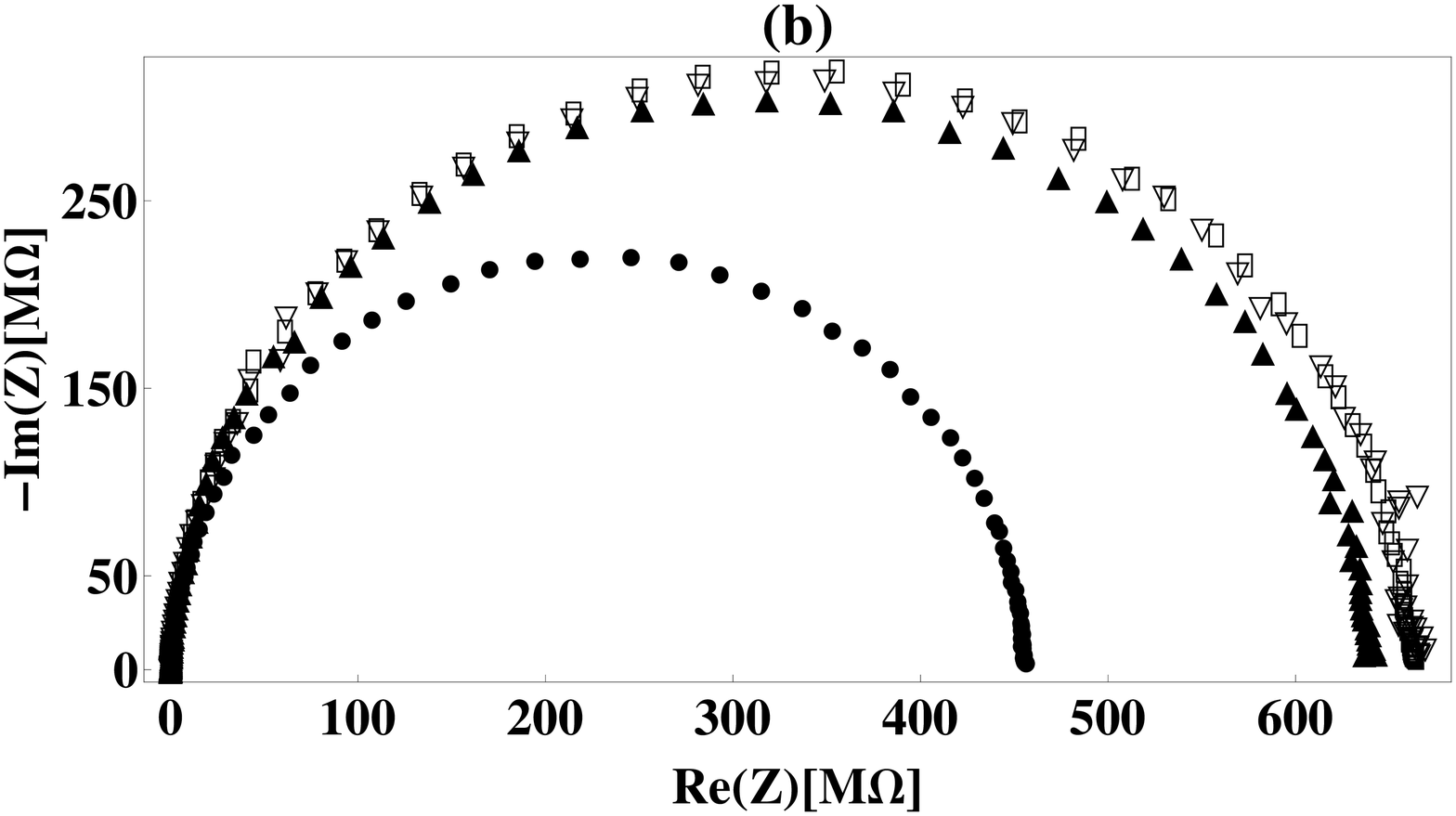}\\
  \includegraphics[width=2.0 in]{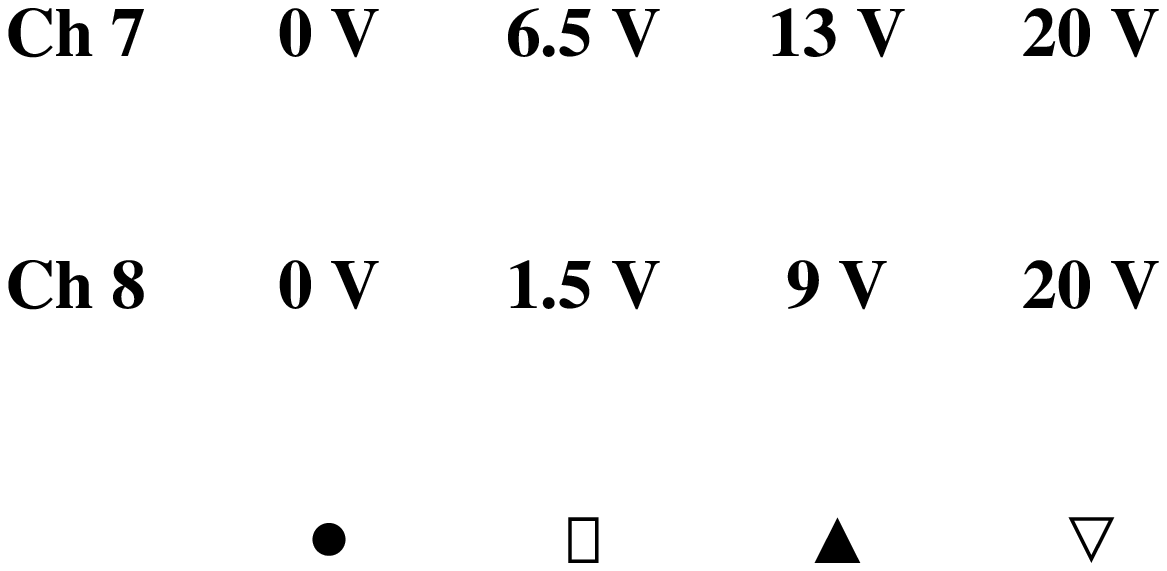}\\
  \end{center}
  \caption{Nyquist plots for channel 7 (a) and 8 (b) showing response at 0V and key points (maxima and minima of resistances). The legend for all figures is shown at the bottom. The hollow rectangles are used to denote the maximum in all cases. As indicated in Fig.~\ref{fig:f4}, the maximum for channel 7 is considerably less well-defined than the other cases.}\label{fig:f7}
\end{figure}
\begin{figure}
   \begin{center}
  \includegraphics[width=3.6 in]{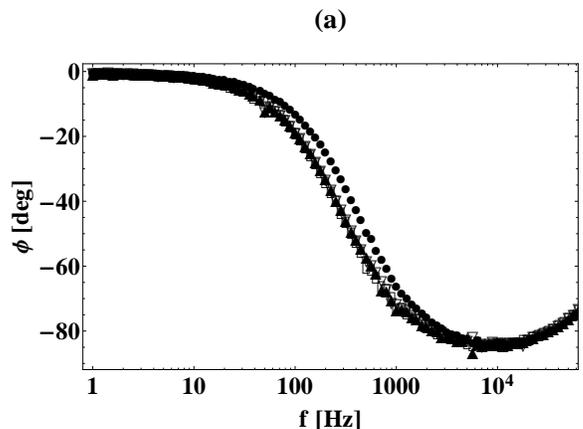}\\
  \includegraphics[width=3.6 in]{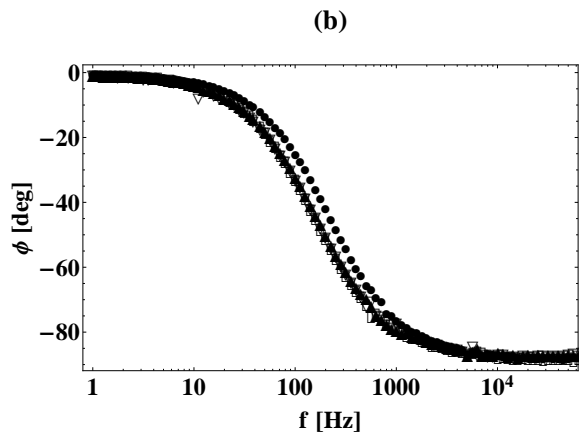}\\
  \end{center}
  \caption{Corresponding phase plots for channel 7 (a) and 8 (b)}\label{fig:f8}
\end{figure}
\renewcommand{\refname}{References}


\begin{thebibliography}{99999}
\bibitem{RZN} I. Rubinstein, B. Zaltzman, A. Futerman, V. Gitis, and V. Nikonenko, Phys. Rev. E \textbf{79}, 021506 (2009)
\bibitem{RZbias} I. Rubinstein, B. Zaltzman, Phys. Rev. E \textbf{80}, 021505 (2009)
\bibitem{Sistat2008} P. Sistat, A. Kozmai, N. Pismenskaya, C. Larchet,
G. Pourcelly, V. Nikonenko, Electrochimica Acta, \textbf{53}, 6380–6390 (2008)
\bibitem{Nikonenko} V.V. Nikonenko and A.E. Kozmai, Electrohimica Acta, \textbf{56}, 1262-1269 (2011)
\bibitem{Schoch} R.B. Schoch, H. van Lintel, and P. Renaud, Phys. Fluids \textbf{17}, 100604 (2005)
\bibitem{DSteinPRL} D. Stein, M. Kruithof, and C. Dekker, Phys. Rev. Lett., \textbf{93}, 035901 (2004)
\bibitem{HeydenPRL} F.H.J. van der Heyden, D. Stein, and C. Dekker, Phys. Rev. Lett. \textbf{95}, 116104 (2005)
\bibitem{KimPRL} S. J. Kim, Y.-C. Wang, J. H. Lee, H. Jang, and J. Han, Phys. Rev. Lett. \textbf{99}, 044501 (2007)
\bibitem{Nanoreview} H.-C. Chang, G. Yossifon, and E. A. Demekhin, Annu. Rev. Fluid Mech. \textbf{44}, 401 (2012)
\bibitem{Naturenano} W. Sparreboom, A. van den Berg, and J. C. T. Eijkel, Nat. Nano \textbf{4}, 713 (2009)
\bibitem{RUBSHTILL79} I. Rubinstein and L. Shtilman, J. Chem. Soc., Faraday Trans. 2 \textbf{75}, 231 (1979)
\bibitem{RZPRE2000} I. Rubinstein and B. Zaltzman, Phys. Rev. E, \textbf{62}, 2238 (2000)
\bibitem{JFM2007} B. Zaltzman and I. Rubinstein, Journal of Fluid Mech., \textbf{579}, 173 (2007)
\bibitem{RZPRL} S.M. Rubinstein, G. Manukyan, A. Staicu, I. Rubinstein, B. Zaltzman, R.G.H. Lammertink, F. Mugele, and M. Wessling, Phys. Rev. Lett, \textbf{101}, 236101 (2008)
\bibitem{YCPRL1} G. Yossifon and H.-C. Chang, Phys. Rev. Lett., \textbf{101}, 254501 (2008)
\bibitem{DydekPRL} E.V. Dydek, B. Zaltzman, I. Rubinstein, D.S. Deng, A. Mani, and M.Z. Bazant, Phys. Rev. Lett, \textbf{107}, 118301 (2011)
\bibitem{KimBazant} S.Nam, I. Cho, J. Heo, G. Lim, M.Z. Bazant, D.J. Moon, G. Sung, S.J. Kim, Phys. Rev. Lett., \textbf{114}, 114501 (2015)
\bibitem{Khoo}J.-H. Han, E. Khoo, P. Bai, and M.Z. Bazant,  Scientific Reports \textbf{4}, 7056 (2014)
\bibitem{DengSuss} D. Deng, E. V. Dydek, J.-H. Han, S. Schlumpberger, A. Mani, B. Zaltzman, and M.Z. Bazant, Langmuir, \textbf{29} 16167-16177 (2013)
\bibitem{Levich} V. G. Levich,  ``\emph{Physico-chemical hydrodynamics,}" Prentice–Hall, (1962)
\bibitem{ManiB}  A. Mani and M.Z. Bazant, Phys. Rev. E, \textbf{84}, 061504 (2011)
\bibitem{Yaroschuk} A. Yaroschuk, E. Zholkoskiy, S. Pogodin, and V. Baulin, Langmuir,\textbf{27},11710-11721 (2011)
\bibitem{NBruus} C.P. Nielsen and H. Bruus, Phys. Rev. E \textbf{90}, 043020 (2014)
\bibitem{Khair} A. Khair, Phys. Fluids, \textbf{23}, 072003 (2011)
\bibitem{ChiaEugene} H.C. Chang, E. A. Demekhin, and V.S. Shelistov, Phys. Rev. E, \textbf{86}, 046319 (2012)
\bibitem{ImpPRL} J. Schiffbauer, S. Park, and G. Yossifon, Phys. Rev. Lett., \textbf{110}, 204504 (2013)
\bibitem{MBonn} D. Lis, E. H. G. Backus, J. Hunger, S. H. Parekh, and M. Bonn, Science, \textbf{344}, 1138-1142 (2014)
\bibitem{surfchgPRE} J. Schiffbauer, U. Liel, and G. Yossifon, Phys. Rev. E, \textbf{89}, 033017 (2014)
\bibitem{EDLthryPRE} J. Schiffbauer and G. Yossifon, Phys. Rev. E, \textbf{89}, 053015 (2014)
\bibitem{YCffPRE} G. Yossifon, P. Mushenheim, Y.-C. Chang, and H.-C. Chang, Phys. Rev. E, \textbf{81}, 046301 (2010)
\bibitem{YoavShachar} Y. Green and G. Yossifon, Phys. Rev. E, \textbf{89}, 013024 (2014)
\bibitem{AbuRjal} R. abu-Rjal, V. Chinaryan, M. Z. Bazant, I. Rubinstein, and B. Zaltzman, Phys. Rev. E, \textbf{89}, 012302 (2014)
\bibitem{BehrensGrier} S.H. Behrens and D.G. Grier, J. Chem. Phys. \textbf{115}, 6716 (2001);




%
%


%
%

%

%

%
%
%
%

%
%
%
%
%
%













\end{thebibliography}
\end{document}